\documentclass[journal,10pt]{IEEEtran}

\usepackage{amsmath}
\usepackage{amsthm}
\usepackage{cite}
\usepackage{color}
\usepackage{graphicx}
\usepackage{epstopdf}
\usepackage{hyperref}
\usepackage{multirow}
\usepackage{steinmetz}
\usepackage{url}
\usepackage[normalem]{ulem}

\begin{document}

	\pagenumbering{gobble}

	\title{Intelligent Time-Adaptive Transient Stability Assessment System}

	\author{James J.Q. Yu, \IEEEmembership{Member,~IEEE},
			David J. Hill, \IEEEmembership{Fellow,~IEEE},
			Albert Y.S. Lam, \IEEEmembership{Senior Member,~IEEE},
			Jiatao Gu, \IEEEmembership{Student Member,~IEEE},
			and Victor O.K. Li, \IEEEmembership{Fellow,~IEEE}

    \thanks{The authors are with the Department of Electrical and Electronic Engineering, The University of Hong Kong, Pokfulam Road, Hong Kong (e-mail: \{jqyu, dhill, ayslam, jiataogu, vli\}@eee.hku.hk).}
}

\markboth{IEEE Transactions on Power Systems}%
{ \MakeLowercase{\textit{Yu}}: Intelligent Time-Adaptive Transient Stability Assessment System}

\maketitle

\begin{abstract}
Online identification of post-contingency transient stability is essential in power system control, as it facilitates the grid operator to decide and coordinate system failure correction control actions.
Utilizing machine learning methods with synchrophasor measurements for transient stability assessment has received much attention recently with the gradual deployment of wide-area protection and control systems.
In this paper, we develop a transient stability assessment system based on the long short-term memory network.
By proposing a temporal self-adaptive scheme, our proposed system aims to balance the trade-off between assessment accuracy and response time, both of which may be crucial in real-world scenarios.
Compared with previous work, the most significant enhancement is that our system learns from the temporal data dependencies of the input data, which contributes to better assessment accuracy.
In addition, the model structure of our system is relatively less complex, speeding up the model training process.
Case studies on three power systems demonstrate the efficacy of the proposed transient stability assessment system.
\end{abstract}


\section{Introduction} \label{sec:intro}

Transient stability is the ability of a power system to retain synchronism subject to large disturbances \cite{Kundur2004}.
It is a significant concern in power system design and operation as it is among the major causes of power blackouts in the past \cite{chiang_direct_2011}.
To meet the ever-increasing growth of power consumption, the system loading introduced in the system causes power grids to operate near the transmission capacities.
This leads to more serious transient stability issues caused by disturbances.
Therefore, real-time assessment of post-disturbance transient stability is still receiving much attention.

Transient stability assessment (TSA) mathematically corresponds to solving a set of high-dimensional non-linear differential algebraic equations (DAE) \cite{fouad_power_1992}.
Transient energy functions have been employed to assess the system stability since 1980s (see  \cite{chiang_direct_2011,fouad_power_1992,Hiskens1989,xue_quantitative_1999} for some references).
However, this approach has its drawbacks when employed on practical large power grids due to the model simplifications required.
An alternative of this direct method is fast time-domain (TD) simulation \cite{kundur_power_1994}, where a given set of credible contingencies are simulated in an off-line manner to guide the design and tuning of the control system.
Meanwhile, accurate TD simulations require complete information of the grid and the disturbance, and impose a heavy computational burden.
These properties hinder TD simulations from being employed for on-line TSA status prediction.
Such assessment results can be significantly different from the actual system response due to the inaccurate system model parameters and disturbance estimations.

Nowadays, the synchrophasor technology is among the key measurement methods in power systems.
Phasor measurement units (PMU) significantly outperform the conventional supervisory control and data acquisition (SCADA) system due to their capability in sampling system variables in synchronism from dispersed locations, enabling system-level model validation \cite{Hashiesh2012}.
With the rapid deployment of PMUs, a collection of methodologies have been proposed for TSA based on the real-time system operating state.
Utilizing the pre- and post-contingency system dynamics, techniques such as piecewise constant-current load equivalent method \cite{Liu1995}, emergency single machine equivalent \cite{Pavella2000}, and post-disturbance trajectory analysis \cite{gurusinghe_post-disturbance_2016} were developed for on-line TSA.
While accurate assessments can be achieved by these methods, their high computational complexities prevent them from being employed in practical post-contingency TSA.

To realize fast real-time TSA status prediction, machine learning and fuzzy logic techniques have been widely adopted as alternative approaches for TSA in recent years, e.g. decision tree methods \cite{Guo2014}, artificial neural networks (ANN)/Support Vector Machine(SVM) \cite{Hashiesh2012,geeganage_application_2015}, and fuzzy knowledge based systems \cite{Kamwa2012}.
Early work \cite{wehenkel_automatic_1998, wehenkel_artificial_1989} was based on decision trees and learning methods.
Different from the conventional analytical methods, these machine learning methods extract the relationship between the system parameters and the corresponding stability conditions utilizing predefined transient stability datasets.
Once such relationships are established, new transient stability cases can be evaluated with minimal computational efforts.
Thus the real-time computing burden of the system is alleviated and the assessment can be conducted in an online manner.

Related work has demonstrated the superiority of these predictive models in constructing TSA systems.
As analyzed in \cite{Kamwa2012}, two competing trends are developed for modern TSA solutions, namely, the fuzzy-logic rule-based approaches and the machine learning-based approaches.
The former has the advantage of transparency, and the internal structures and parameters have physical meanings while the latter has generally better performance.
This  leads to a trade-off between TSA transparency and accuracy.
Meanwhile, since it is possible for the instability to propagate over the network within seconds, remedial actions are usually automated.
In such cases, as analyzed in \cite{Kamwa2012}, assessment accuracy is critical for TSA prediction, and machine learning-based techniques are employed in this work to construct an intelligent TSA prediction system.
Note that we do not intend to compare the usefulness of the two main trends.
In power applications where a transparent model is necessary, fuzzy-logic-based approaches may be preferable \cite{Kamwa2012}.

However, there is still one major research gap in machine learning-based TSA techniques.
Existing work tends to employ a fixed response time scheme for implementation.
The proposed TSA systems have to wait for a fixed period of time before evaluating the stability status of the grid.
As the dimension and type of input data have been defined, such implementations can drastically decrease the complexity of system design.
However, as transient stability issues can quickly propagate over the whole grid, it is always preferred to have a faster TSA response so that more time can be reserved for remedial control actions to take effect.
Moreover, it is possible to identify the issues at the very early stage after fault clearance, and the existing approaches with fixed response time cannot dynamically adapt to such cases.
Therefore, it is more advantageous to conduct TSA in an adaptive online manner.
A recent work in \cite{Zhang2015a} reported a novel methodology to make transient stability assessment at the earliest time.
However, as multiple decision making machines are trained separately for measurements with different timestamps, the training process discards the potential causal relationship of one measurement and its descendant ones.

In order to utilize this information for TSA prediction, in this work we propose a long short-term memory (LSTM) \cite{Hochreiter1997} based method.
The main contributions of this work are as follows:
\begin{itemize}
\item This work is the first attempt to employ recurrent neural network (RNN) and LSTM for a time-adaptive TSA process.
The proposed system manages to extract both spatial and temporal data dependency from the input power system state for security assessment, which is novel in stability analysis methodologies.
\item The system framework is significantly easier to implement than previous time-adaptive TSA \cite{Zhang2015a}, resulting in reduced training time and maintenance effort.
\item The proposed system is assessed against three small to large scale power system test cases.
The simulation results demonstrate accurate assessment with significantly reduced system response time.
\item Parameter sensitivity tests are carried out to evaluate the performance of the system.
\end{itemize}

The remainder of this paper is organized as follows.
Section II briefly introduces the LSTM network employed in this work.
Section III elaborates our proposed LSTM-based TSA System (TSAS), and Section IV develops a scheme to make time-adaptive assessments.
Section V demonstrates numerical results on the benchmark test systems with parameter sensitivity study.
Finally, we conclude this work in Section VI.

\section{Transient Stability Assessment and Recurrent Neural Networks}

In this section we first introduce TSA, and then give a brief overview on RNN and LSTM networks.
Such networks are critical in the proposed TSAS which will  be presented in Section \ref{sec:proposed}.

\subsection{Transient Stability Assessment}

With the gradual deployment of synchrophasor measurement devices, wide area protection and control (WAPaC) systems have attracted significant attention \cite{begovic_wide-area_2005}.
To implement such a response-based protection system, TSA needs to be conducted after the clearance of system faults to predict the system stability, i.e., whether the system can maintain its synchronism subject to the disturbance and operating condition.

\begin{figure}
	\centering
	\includegraphics[width=\linewidth]{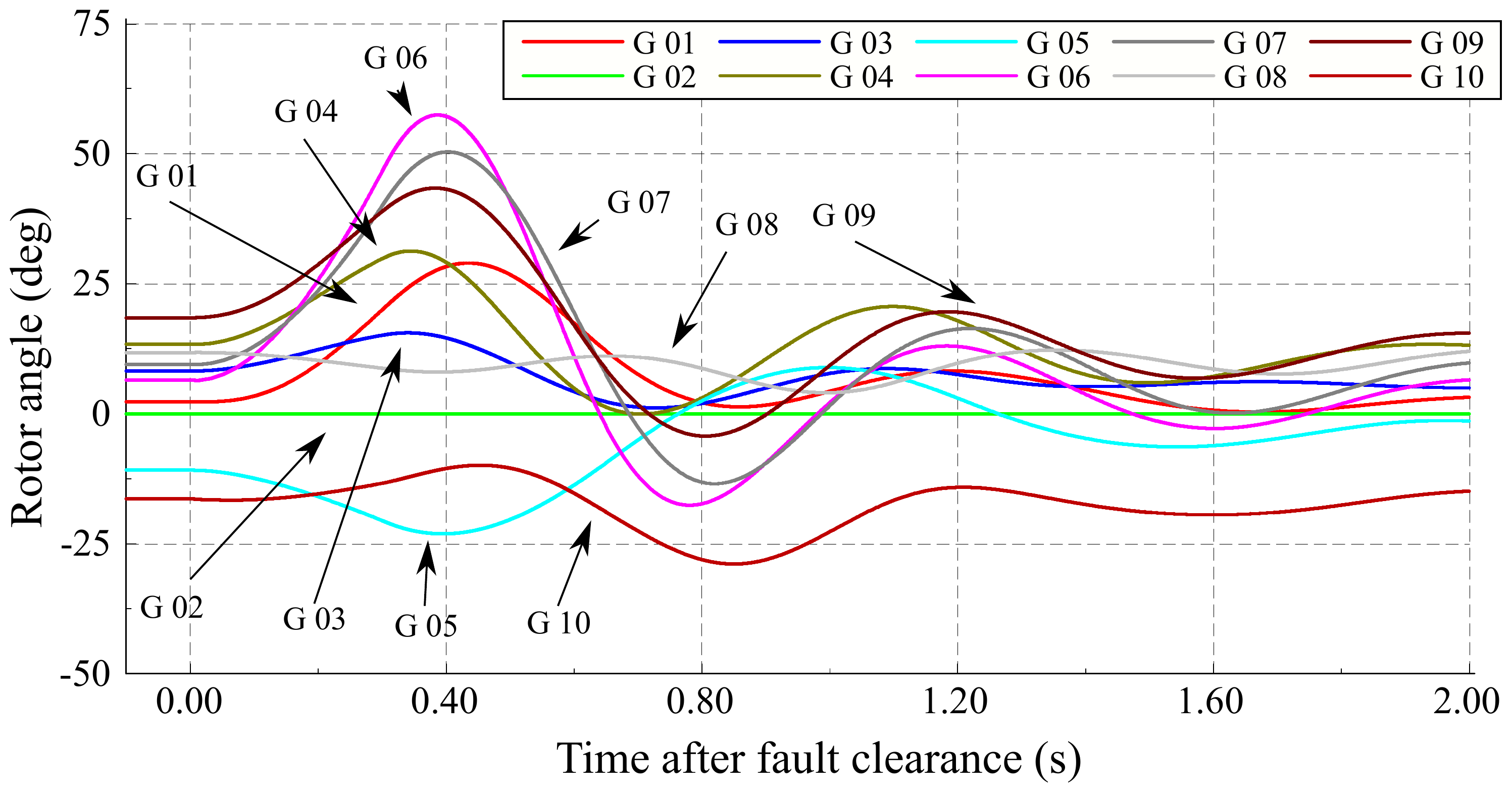}
	\caption{Synchronous generator rotor angle dynamics after a system fault in New England 10-machine system \cite{Pai1989}.}
	\label{fig:TSA}
\end{figure}

Fig. \ref{fig:TSA} depicts the generator rotor angle dynamics of a system fault in the New England 10-machine system, where bus 16 experienced a three-phase short circuit, and the fault is cleared after 0.25 second. In this figure, the number after ``G'' in the legend indicates the index of the corresponding generator. Generator 2 is used as the reference machine.
Each line in the plot represents the rotor angle dynamics of one synchronous machine.
In such transient situations, it is desired that the system behavior is observed, and the future stability status of the system is predicted in the earliest possible time.
If the system will lose its synchronism, emergency control actions will take place to maintain stability.
This prediction process is called TSA, and a faster system response speed is always preferred to provide extra time for control actions.
Here TSA accuracy and response speed form a trade-off, and it is optimal to evaluate the trade-off based on the characteristics of contingencies.
We call the schemes with this capability \textit{time-adaptive TSA} techniques, as they can adapt to different contingencies and generate the accurate assessment results at the earliest possible time.

\begin{figure}
	\centering
	\includegraphics[width=\linewidth]{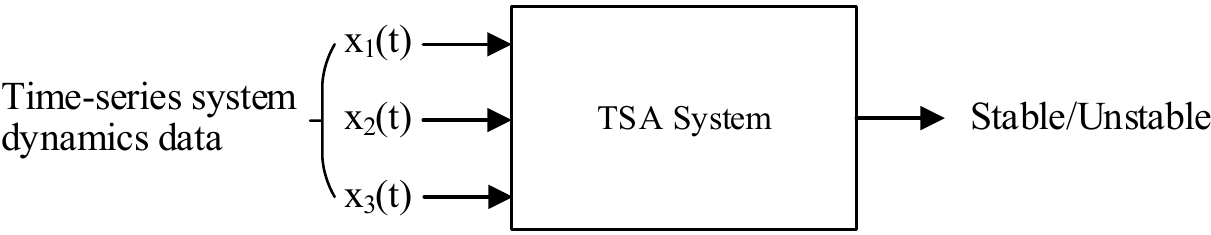}
	\caption{Typical design of machine learning TSA systems.}
	\label{fig:TSA-System}
\end{figure}

To design such time-adaptive methods, we investigate the components that constitutes the response time.
Modern fast TSA based on machine learning techniques can be generalized in Fig. \ref{fig:TSA-System}.
The post-disturbance system dynamics are digitized into time-series system dynamic information, which is input into the TSA system for developing a future system stability index.
It is clear that the response time is mostly influenced by two factors: data aggregation time and TSA system computation time.
Therefore, in this work we propose a time-adaptive scheme to minimize the volume of data required, and construct a computationally efficient system to handle the calculation.

\subsection{Recurrent Neural Network and Long Short-Term Memory}

RNN \cite{Narendra1990} is a special type of neural network and it considers data correlation in the time domain.
A typical RNN takes a temporal sequence of vectors  $[\mathbf{x}_1, \mathbf{x}_2, \cdots, \mathbf{x}_T]$ as input, and outputs a sequence of vectors $[\mathbf{h}_1, \mathbf{h}_2, \cdots, \mathbf{h}_T]$.
The output is generated by the following equation for $t=1,2,\cdots,T$:
\begin{equation}
\mathbf{h}_t=f(\mathbf{W}\mathbf{x}_t + \mathbf{H}\mathbf{h}_{t-1}+\mathbf{b}),
\end{equation}
where $f(\cdot)$ is a non-linear activation function, and $\mathbf{W}$, $\mathbf{H}$, $\mathbf{b}$ are the learning parameters.
Different from ANNs, RNNs have additional recurrent connections in the hidden layers, which provide essential memory capabilities.
Such connections facilitate the network to keep previous information for later use in the time domain, and thus capture the time-domain dependencies in the input data.

While the RNN architecture utilizes the data correlation information, typical gradient-based training algorithms suffer from deteriorated memory performance as the long-range data correlation is undermined \cite{Bengio1994}.
They suffer from the ``vanishing gradient problem'' \cite{Hochreiter1991}, resulting in poor performance in capturing long temporal dependencies.
As an alternative, a novel RNN architecture was proposed in \cite{Hochreiter1997} to address this problem.
This improved model is referred to as long short-term memory (LSTM).

\begin{figure}
	\centering
	\includegraphics[width=\linewidth]{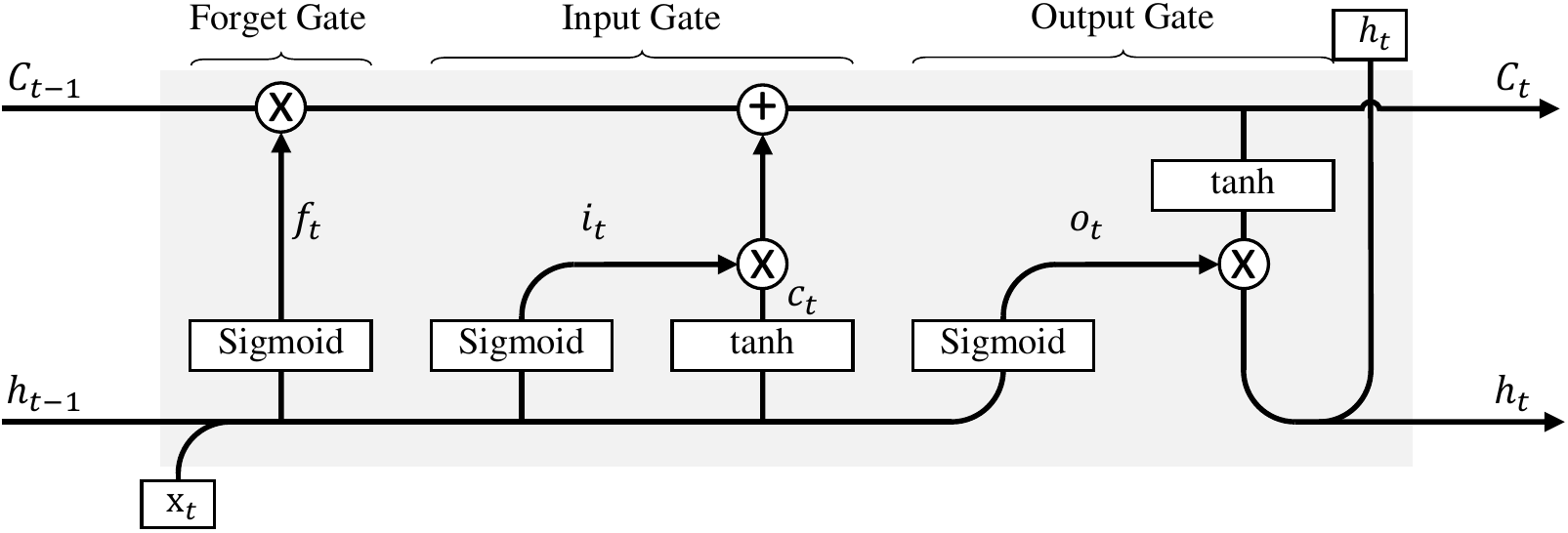}
	\caption{An illustration of LSTM memory cell \cite{UnderstandingLSTMNetworks}.}
	\label{fig:LSTM}
\end{figure}

LSTM, designed to overcome the long temporal dependency defect in RNN, is a kind of network that implements memory blocks, containing one or more memory cells \cite{Hochreiter1997}.
These cells are employed to maintain a long term memory (state) over time when combined with a typical RNN.
The architecture of LSTM is presented in Fig. \ref{fig:LSTM}.

\begin{figure}
	\centering
	\includegraphics[width=0.7\linewidth]{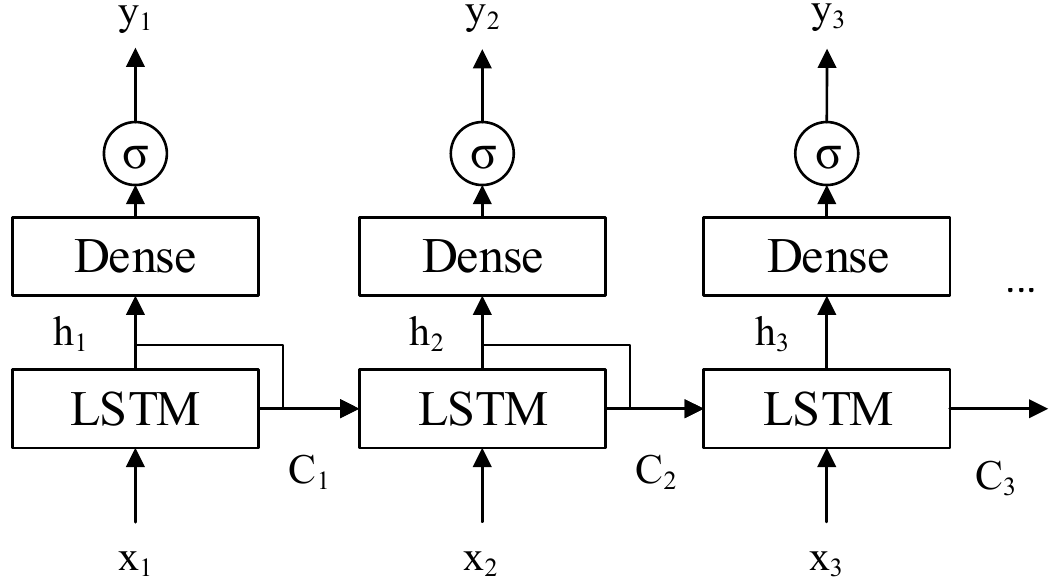}
	\caption{Structure of TSAS. Given a sequence of $x_t$ data, the system can develop corresponding $y_t$ predictions.}
	\label{fig:Structure}
\end{figure}

As illustrated in Fig. \ref{fig:LSTM}, LSTM comprises three gates: input, forget, and output gates.
These gates cooperate with a memory cell $\mathbf{C}_t$ to maintain the LSTM cell state:
\begin{IEEEeqnarray}{rCl} \IEEEyesnumber \label{eqn:input_forget_gate}
\mathbf{i}_t &=& \sigma(\mathbf{W}_i\mathbf{x}_t+\mathbf{U}_i\mathbf{h}_{t-1}+\mathbf{b}_i) \IEEEyessubnumber\\
\mathbf{c}_t &=& \tanh(\mathbf{W}_C\mathbf{x}_t+\mathbf{U}_C\mathbf{h}_{t-1}+\mathbf{b}_C) \IEEEyessubnumber\\
\mathbf{f}_t &=& \sigma(\mathbf{W}_f\mathbf{x}_t+\mathbf{U}_f\mathbf{h}_{t-1}+\mathbf{b}_f) \IEEEyessubnumber\\
\mathbf{C}_t &=& \mathbf{f}_t\ast \mathbf{C}_{t-1} + \mathbf{i}_t\ast \mathbf{c}_t, \IEEEyessubnumber
\end{IEEEeqnarray}
and update the output representation:
\begin{IEEEeqnarray}{rCl} \IEEEyesnumber \label{eqn:output_gate}
\mathbf{o}_t &=& \sigma(\mathbf{W}_o\mathbf{x}_t+\mathbf{U}_o\mathbf{h}_{t-1}+\mathbf{b}_o) \IEEEyessubnumber\\
\mathbf{h}_t &=& \mathbf{o}_t\ast \tanh(\mathbf{C}_t), \IEEEyessubnumber
\end{IEEEeqnarray}
where $\ast$ is the element-wise product, $\sigma(x)=\frac{1}{1+e^{-x}}$ is a sigmoid function, and $\mathbf{W}$, $\mathbf{U}$, $\mathbf{b}$ are matrices corresponding to the learning parameters $\mathcal{P}=[\mathbf{W},\mathbf{U},\mathbf{b}]$.
Numerous LSTM-based applications have been developed due to its simplicity and efficacy.
Interested readers can refer to \cite{lecun_deep_2015} for a thorough introduction of the methodology.

\section{LSTM-based TSAS} \label{sec:proposed}

The LSTM-based TSAS proposed in this paper is based on the hypothesis that the post-contingency power system measurements, e.g., bus voltage phasors, can immediately indicate the stability of the system after experiencing the disturbance, see \cite{wehenkel_artificial_1989, gomez_support_2011} for some references.
In addition, the temporal data dependency of the measurements can reduce the data quantity required for accurate assessments.
Based on these hypotheses, we establish a non-linear relationship between the observed measurements and the system stability by using LSTM as the primary classifier.
The structure of the learning system is presented in Figs. \ref{fig:Structure}.

In the proposed TSAS, two layers of artificial neurons are employed to extract the characteristics of the input data, namely, an LSTM memory block layer and a hidden dense neuron laye.
The LSTM layer is utilized to handle input power system measurements, and outputs  $\mathbf{h}_t$, sharing the same dimensionality of $\mathbf{x}_t$.
As the extracted data feature by the LSTM layer is not human-readable, an additional dense hidden layer is appended for dimensionality reduction.
In addition, a final $\mathrm{Sigmoid}$ function is employed to normalize the output solution.

\subsection{Time-Series Simulation}

To enable the proposed TSAS for real-time TSA, it needs to be trained offline.
The training data is generated using time-series simulation of different transient contingencies on the given power system.
In this system, we employ the positive sequence voltage phasors of all buses as input data, which in real-time can be measured with synchophasor measurement techniques.
Using 50/60-Hz sampling of the simulated measurements, the measured voltage phasor data of an arbitrary contingency after fault clearance is in the form of
\[
\begin{bmatrix}
V_{1,1} & V_{1,2} & \cdots & V_{1,T} \\
V_{2,1} & V_{2,2} & \cdots & V_{2,T} \\
\vdots & \vdots & \ddots & \vdots \\
V_{B,1} & V_{B,2} & \cdots & V_{B,T}
\end{bmatrix},
\]
where $B$ and $T$ are the total number of buses and the length of the post-contingency observation window, respectively. $V_{b,t}=\mathrm{V}_{b,t}\phase{\mathrm{\theta}_{b,t}}$ is the voltage phasor of bus $b$ at time instance $t$.
While $B$ is determined by the system topology, $T$ is a control parameter influencing the LSTM complexity and assessment accuracy; a large $T$ will result in a high nonlinearity of the TSAS, while a small $T$ may impair the input data completeness, which in return weakens the accuracy.

In addition, the synchronous machine rotor angle is also calculated in the time-series simulation, and the maximum angle deviation of any two machines at any time is recorded, denoted by $\delta_\text{max}$.
This value is later utilized to determine the system stability after contingencies.

\subsection{Offline Training}

To better simulate the system response after various disturbances, a large set of defined contingencies are utilized in the time-series simulation.
Employing the voltage phasor and maximum angle deviation data for each training contingency, the training dataset can be established where the input of each training case is presented as $\mathbf{x}=[\mathbf{x}_1, \mathbf{x}_2, \cdots, \mathbf{x}_T]$, where
\begin{equation}
\mathbf{x}_t=[\mathrm{V}_{1,t},\mathrm{V}_{2,t},\cdots,\mathrm{V}_{B,t}, \phase{\mathrm{\theta}_{1,t}}, \phase{\mathrm{\theta}_{2,t}},\cdots,\phase{\mathrm{\theta}_{B,t}}]^\top.
\end{equation}

Meanwhile, to construct a complete training case, the output assessment result $\mathbf{y}$ is obtained by observing $\delta_\text{max}$:
\begin{equation}\label{eqn:y_value}
y_t = \begin{cases}
1\text{ (Stable)} & \text{for }\eta>0\\
0\text{ (Unstable)} & \text{for }\eta\leq0\\
\end{cases},t = 1,2,\cdots,T,
\end{equation}
where $\eta = (360-\delta_\text{max})/(360+\delta_\text{max})$,	and $\delta_\text{max}$ is the maximum angle deviation of any two generators in the LSTM system at any time. $\eta$ is a power angle-based stability index \cite{Pavella2000} in which the generator rotor angle difference $\delta$ is used  to determine if any generator in the system is out of synchronism.
This index is widely employed as a transient stability index in the previous literature, see \cite{Zhang2015a,gomez_support_2011} for some references.

Given a collection of $N$ training cases $\{\mathbf{x}_{(n)},y_{(n)}\}_{n=1}^N$, the objective of the offline training is to obtain the system parameters $\mathbf{W}$, $\mathbf{U}$, and $\mathbf{b}$.
In this paper, the Adam optimizer \cite{DiederikKingma2015} is employed to find the optimal values of these parameters with the binary cross entropy error function as the objective:

\begin{equation}
\text{minimize}\quad -\sum_{n=1}^N[y_{(n)}\log\hat{y}_{(n)}+(1-y_{(n)})\log(1-\hat{y}_{(n)})],
\end{equation}
where $\hat{y}_{(n)}$ is the actual classification (assessment) result of $\mathbf{x}_{(n)}$ on the trained system.

\subsection{Online Assessment}

For online assessment of the transient stability status, the previously trained TSAS is employed to classify the test cases using post-contingency PMU measurements.
With a set of bus voltage phasor measurements of length $T$, it is trivial to calculate the corresponding $\mathbf{h}_T$ and thus $\hat{y}_T\in(0, 1)$ using the trained system parameters.
In order to transform the system output $\hat{y}_T\in(0, 1)$ into a stability index, a bipartite threshold $\delta=0.5$ is defined such that the test cases with $\hat{y}_T<\delta$ is considered unstable, and otherwise stable.

\begin{figure*}
	\centering
	\includegraphics[width=0.75\linewidth]{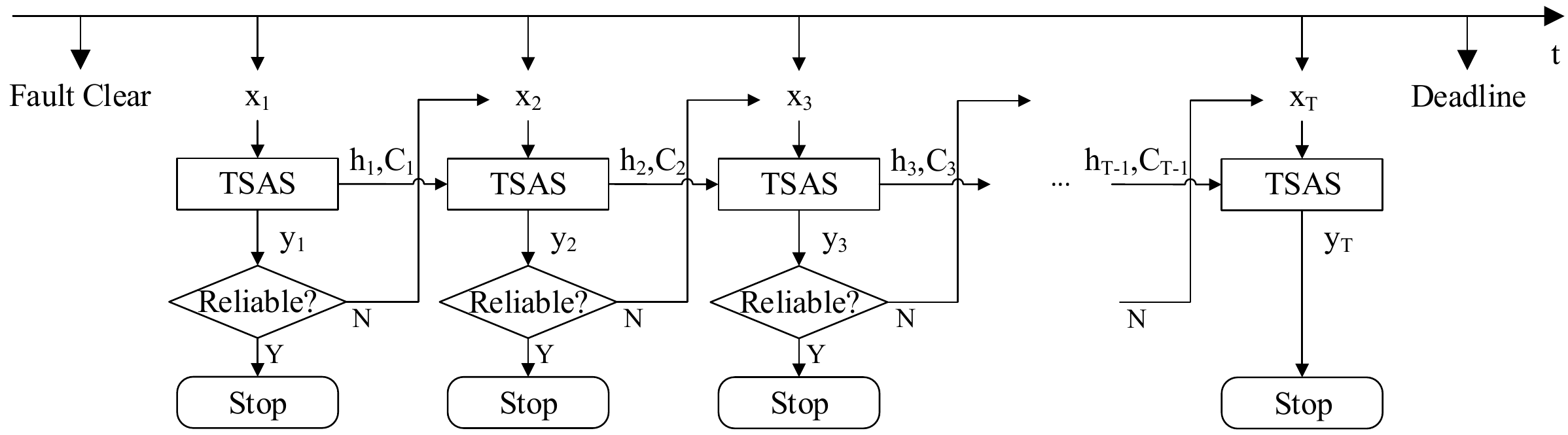}
	\caption{Flow chart of time-adaptive TSAS.}
	\label{fig:Time-Adaptive}
\end{figure*}

\section{Time-adaptive TSAS}

\subsection{Time Adaptive Implementation}

Similar to most existing TSA models, the method proposed in Section \ref{sec:proposed} utilizes a fixed-length observation window, i.e., the response time is constant.
However, this static time response can be less efficient when dealing with fast transient instabilities.
Moreover, different system models may require substantially different lengths of observation windows to obtain reliable assessment conclusions.

In this section, we propose a time-adaptive TSAS based on the LSTM system in Section \ref{sec:proposed}.
The objective of the time-adaptive scheme is to generate a reliable assessment result as fast as possible, which is a trade-off between the assessment speed and accuracy.
Consequently, control actions can take place at an early time to avoid possible system failures.

Contributed by the recurrent connections of LSTM memory blocks, our proposed TSAS takes the temporal data dependency into consideration when assessing the system stability.
However, as only $\hat{y}_T$ is concerned in generating stability indices, potential information loss happens when $\hat{y}_t, t=1,2,\cdots,T-1$ values are discarded.
Utilizing this information, we propose a time-adaptive TSAS for fast stability assessment.

The flow chart is presented in Fig \ref{fig:Time-Adaptive}, where all ``TSAS'' blocks are identical with constant trained system parameters.
After the fault clearance, TSA will be conducted immediately upon the receipt of PMU measurements at the first time instance.
As $\mathbf{x}_1$ is available at the system, $\hat{y}_1$ can be calculated and utilized to determine the transient stability index.
Consequently, the system stability can be assessed as follows, where (\ref{eqn:y_value}) is transformed to determine the reliability:
\begin{equation}\label{eqn:y_value_modified}
\text{Stability} = \begin{cases}
\text{Stable} & \text{for }\hat{y}_t>1-\delta\\
\text{Unstable} & \text{for }\hat{y}_t<\delta\\
\text{Unknown} & \text{otherwise }\\
\end{cases},t = 1,2,\cdots,T.
\end{equation}
In this equation, the stability threshold $\delta\in(0,0.5)$ is used to manipulate the trade-off between speed and accuracy.
A larger $\delta$ may result in an earlier assessment but the accuracy may be undermined, while a smaller $\delta$ will lead to potentially more accurate result at the sacrifice of assessment speed.

If (\ref{eqn:y_value_modified}) returns ``Stable'' or ``Unstable'', the system outputs this conclusion for further control operations.
Otherwise, the system will wait for the next time instance, and the newly received measurements $\mathbf{x}_t$ will be included in the calculation of the corresponding $y_t$ values.
This process repeats until a reliable result is achieved, or the maximum decision-making time $T^\textit{max}$ is reached.

The proposed TSAS relates to the system in \cite{Zhang2015a} in the following sense.
Both systems can perform online TSA in a time-adaptive manner.
However, due to the difference in the underlying methods utilized (LSTM versus ELM), our proposed TSAS extracts additional temporal data-dependency features over the approach in \cite{Zhang2015a}.
Moreover, as we employ only one LSTM network in the proposed TSAS, the training complexity is greatly reduced when compared with the multi-ELM system in \cite{Zhang2015a}.
This also yields a simpler yet still accurate decision making rule (\ref{eqn:y_value_modified}), whose performance will be demonstrated in the following sections.

\subsection{Discussion}

In the time-adaptive intelligent system, LSTM plays a key role in predictions based on available time-series data.
As a variant of RNN and ANN, LSTM has two major advantages that make it suitable for TSA:
\begin{itemize}
	\item The most computationally expensive process, i.e., training, can be performed in an offline manner.
	The online TSA testing process can be run very effectively given the trained neural network parameters \cite{hansen_neural_1990}.
	\item Neural networks can model transient stability in power networks with power dynamics data \cite{narendra_identification_1990}. This allows us to relax the requirement on accurate knowledge of power system models in practice.
\end{itemize}
Hence the problems (requirement of complete power system model and heavy real-time computational burden as discussed in Section \ref{sec:intro}) encountered by the traditional non-artificial-intelligence TSA techniques can now be overcome with the proposed system.

Meanwhile, neural networks are often regarded as blackbox models because the internal network parameters and variables generally do not  carry explicit physical meanings.
However, the lack of these physical meanings does not influence the practicality of ANN, since only the output value is required for subsequent analyses and/or calculations.
As the interior structure can be defined by \eqref{eqn:input_forget_gate} and \eqref{eqn:output_gate}, only algebraic calculations are needed to generate the output from the input power system dynamics sampled by PMUs.
In the case of TSAS, the output of the neural network holds the transient stability status of the system.

\section{Case Studies}

The New England 10-machine system \cite{Pai1989} is adopted to evaluate the assessment performance of the proposed TSAS.
All numerical simulations are conducted on a computer with an Intel Core i7 CPU working at 3.4 GHz and 8 GB RAM.
The LSTM framework is constructed with Theano \cite{Theano} and Keras \cite{Keras}.
The proposed system is implemented in Python.

For TSA using supervised learning algorithms, sufficient previous knowledge that depicts the post-contingency characteristics of the grid is critical \cite{Xu2011}.
The training cases should include adequate information to guide the learning of TSAS to approximate the system behavior under different operation conditions.

The training and testing cases are generated by time-domain simulation of post-contingency power system dynamics.
In the simulation, we consider both the nominal power network topology and the $(N-1)$ contingencies, in which any one of the transmission lines or transformers is out of service.
The consumed power is set to 80\%, 100\%, and 120\% of the basic system load levels, respectively\footnote{For different system load levels, both generations and loads are scaled by the ratio.}.
For each load level, three-phase short-circuit faults are assumed to occur on either buses, or transmission lines where the faults are located at 20\%, 40\%, 60\%, and 80\% of the whole line \cite{Zhu2015}.
Further, the fault clearance time is set to arbitrary values between 0.1s and 0.4s.
Consequently, 5000 TSA contingency cases are generated with the above configurations.
The number of cases is similar to that of the same network in previous work \cite{Zhang2015a}.

For each contingency case, the time-domain simulation is conducted using TSAT \cite{DSATools}, and the calculated bus voltage phasors are employed as the measurements sampled by PMUs.
In practice, utilities can utilize the historical operation data to train TSAS (see \cite{Kamwa2012} for an example).
By this means the extensive simulation time for generating power system dynamics under these contingencies can be avoided.
Note that unless specified, we assume that PMU can accurately sample power system variables.
The impact of inaccurate PMU measurements will be investigated in Section \ref{sub:noisy}.

For cross validating and preventing the over-fitting problem, in this work we randomly divide the generated contingency cases into training and testing sets in the ratio of 3:1, which accords with  \cite{Zhang2015a,Zhu2015}.
Only the training cases are employed to train the TSAS. The testing cases are used to assess the TSA accuracy when TSAS encounters unknown contingencies.
Thus over-fitting problem (which describes the models learning from both characteristics and random noises in the input data, leading to poor predicting performance) can be avoided.
Generally over-fitted models will result in inferior accuracy for testing cases, which is not observed in our simulations.

In order to train the proposed TSAS, control parameters shall be defined.
In this work, two layers of LSTM memory blocks are employed in the LSTM layer, where each layer has 128 memory blocks.
The numbers of input and output nodes are $2B$ and $1$, respectively.
The network is trained for up to 100 epochs, subject to an early stopping method to prevent overfitting.
In the tests, the maximum decision-making time $T^\textit{max}$is set to $20$.
Parameters $\delta$ and $T$ are set to $0.4$ and $5$, respectively, and their sensitivities will be studied later.

\subsection{TSA Accuracy and Response Time}

\begin{table*}
	\centering
	\caption{TSA Accuracy and Average Response Time on New England 10-machine Test System}
	\label{tbl:10-machine}
	\begin{tabular}{r|rrrr|rrrr}
	\hline
	\multirow{2}*{$t$} & \multicolumn{4}{c|}{Training Set} & \multicolumn{4}{c}{Testing Set} \\
	\cline{2-9}
	& Unknown & Correct & Wrong & Accuracy & Unknown & Correct & Wrong & Accuracy \\
	\hline
	0 & 3750 & 0 & 0 & N/A & 1250 & 0 & 0 & N/A \\
	1 & 1103 & 2647 & 0 & 100.00\% & 410 & 840 & 0 & 100.00\% \\
	2 & 176 & 3574 & 0 & 100.00\% & 54 & 1196 & 0 & 100.00\% \\
	3 & 105 & 3645 & 0 & 100.00\% & 42 & 1208 & 0 & 100.00\% \\
	4 & 72 & 3678 & 0 & 100.00\% & 21 & 1229 & 0 & 100.00\% \\
	5 & 58 & 3692 & 0 & 100.00\% & 21 & 1229 & 0 & 100.00\% \\
	6 & 29 & 3721 & 0 & 100.00\% & 12 & 1238 & 0 & 100.00\% \\
	7 & 3 & 3747 & 0 & 100.00\% & 0 & 1250 & 0 & 100.00\% \\
	8 & 0 & 3749 & 1 & 99.97\% & 0 & 1250 & 0 & 100.00\% \\
	\hline
	\end{tabular}
\end{table*}

Table \ref{tbl:10-machine} presents the TSA test results for the New England 10-machine system.
The assessment result is presented for both training cases and testing cases for completeness.
In this table, the values in the ``Unknown'' column represent the total number of test instances whose stability indices cannot be computed at time $t$ after fault clearance.
The simulation results can lead to a conclusion that the majority of the contingencies can be correctly assessed at a very early stage after the clearance of contingencies.
For both training and testing cases, around two thirds of all instances can be assessed at the first post-contingency cycle.
In addition, the early assessment accuracy is perfect for both training and testing cases.
This outstanding performance is potentially contributed by the deeper neural network structures compared with the existing TSA with ANN work \cite{Hashiesh2012, Zhang2015a}.
Our proposed TSAS with one layer of LSTM is also tested under the same simulation environment and input dataset for comparison, and the assessments are less than 99\% accurate from the first post-contingency cycle.

After the first post-contingency cycle, the assessment accuracy to determine those originally ``Unknown'' cases remains perfect for testing cases when given more measurements for assessment, and there is only one wrong assessment among all 3750 training cases.
Moreover, only around 4\% of the contingency cases cannot be assessed with data of two cycles, and all instances can be correctly assessed within seven post-contingency cycles for testing cases, and eight for training cases.

Compared with the other time-adaptive approach given in \cite{Zhang2015a}, it can be observed that our proposed TSAS can maintain a superior accuracy when dealing with the test instances that require more information for classification.
Intuitively the performance improvement originates from the employed LSTM network, which extracts the data features from the temporal dependencies of the training data.

In addition to the assessment accuracy, the response speed is another major concern when evaluating the performance of TSA methods.
It can be easily derived from Table \ref{tbl:10-machine} that the average response time (ART) \cite{Zhang2015a} of TSAS on the New England 10-machine system is 1.448 cycles with 100\% accuracy for testing cases, and 1.412 cycles with 99.97\% accuracy for training cases.

To see the performance improvement of TSAS over the existing TSA techniques, we compare the ART and accuracy of the methodologies proposed in the literature with TSAS in Table \ref{tbl:comparison}.
In this table, the simulation performance\footnote{For fairness of comparison, the simulation results for testing cases are demonstrated as did in the previous literature \cite{Zhang2015a}.} of the proposed TSAS on the New England 10-machine system and two large scale test systems to be introduced in Section \ref{sub:large} are listed.
Meanwhile, as the response time of the other time-adaptive methods can be significantly affected by the power system size, the simulation results for test systems of different scales are presented.
For other non-time-adaptive methods, the best assessment accuracies are compared.

From the results, it can be concluded that the proposed TSAS can outperform existing state-of-the-art TSA techniques.
TSAS achieves a similar ART performance compared with the ELM-based system proposed in \cite{Zhang2015a} (1.448 cycles vs 1.4 cycles for 10-machine system, 2.047 cycles vs 2.8 cycles for 50-machine system), and provides better assessment accuracy in both of the test systems.
While the SVM-based system \cite{gomez_support_2011} and Prediction-based system\cite{Rajapakse2010} give accurate assessment results for all test cases, TSAS achieves the same performance with much shorter time, which consequently allows us to reserve more time for subsequent control actions against unstable situations.

\begin{table}
\centering
\caption{Comparison with previous TSA methodologies}
\label{tbl:comparison}
\begin{tabular}{l|rr}
	\hline
	Method & ART & Accuracy \\
	\hline
  Proposed TSAS on 10-machine & 1.448 cycles & 100.00\% \\
  Proposed TSAS on 162-bus (17-machine) & 1.901 cycles & 100.00\% \\
  Proposed TSAS on 145-bus (50-machine) & 2.047 cycles & 99.98\% \\
	\hline
  ELM-based system on 10-machine \cite{Zhang2015a} & 1.4 cycles & 99.10\% \\
  ELM-based system on 50-machine \cite{Zhang2015a} & 2.8 cycles & 99.70\% \\
  SVM-based system \cite{gomez_support_2011} & 4 cycles & 100.00\% \\
  Prediction-based system \cite{Rajapakse2010} & 6 cycles & 100.00\% \\
  DTFR-based system \cite{kamwa_development_2009} & 1--2 seconds & 95.00\% \\
	\hline
\end{tabular}
\end{table}

\begin{figure}
	\centering
	\includegraphics[width=\linewidth]{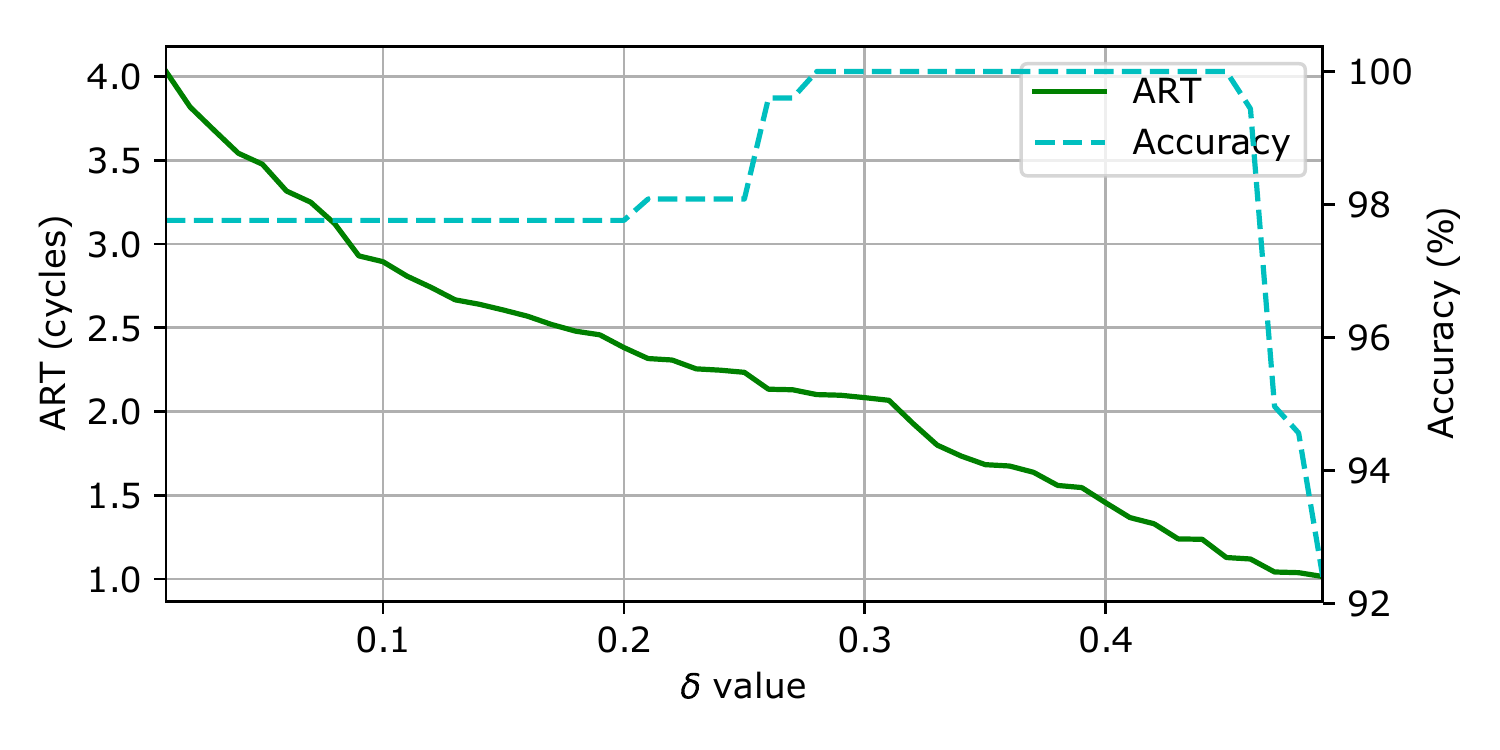}
	\caption{$\delta$ sensitivity test on New England 10-machine system.}
	\label{fig:delta_sensitivity}
\end{figure}

\begin{table}
	\centering
	\caption{$T$ Sensitivity Test on New England 10-machine Test System}
	\label{tbl:T_sensitivity}
	\begin{tabular}{l|rrrrr}
		\hline
		$T$ & 1 & 2 & 3 & 4 & 5 \\\hline
		ART (cycles) & 1.009 & 1.095 & 1.209 & 1.342 & 1.455 \\
		Accuracy & 95.28\% & 98.08\% & 98.80\% & 98.88\% & 100.00\% \\
		\hline\hline
		$T$ &  6 & 7 & 8 & 9 & 10\\\hline
		ART (cycles) &  1.743 & 1.819 & 2.162 & 1.962 & 2.231 \\
		Accuracy & 99.92\% & 100.00\% & 100.00\% & 100.00\% & 100.00\% \\
		\hline
	\end{tabular}
\end{table}

\subsection{$\delta$ and $T$ Sensitivity Tests}

In the proposed TSAS, two system parameters significantly influence the assessment performance, namely the stability threshold $\delta$ and training observation window length $T$.
Two sets of simulations are performed to evaluate their sensitivities, and the results are depicted in Fig. \ref{fig:delta_sensitivity} and Table \ref{tbl:T_sensitivity}.
In Fig. \ref{fig:delta_sensitivity}, 49 $\delta$ values in the range [0.1,0.49] with a step size of 0.01 are simulated, and the ART and accuracy results are plotted.
While a smaller ART is preferred, the system prediction accuracy is always an important performance metric in TSA.
This figure indicates that $\delta$ values around 0.4 are preferred in order to achieve a perfect assessment accuracy while maintaining a fast response time.

The sensitivity of the training observation window length $T$ is presented in Table \ref{tbl:T_sensitivity}.
For example, $T=5$ means that the system is trained using the voltage phasors of the first five cycles after fault clearance, and tested with continuous time-series data.
Therefore, this test aims to identify the influence of input data size on the system performance.
From the table it can be observed that while the ART performance decreases with the increase of $T$, the assessment accuracy demonstrates a reverse trend.
While a small $T$ can lead to a fast response time, a medium $T$ value, e.g. 5, is the optimal choice of the parameter considering the trade-off between ART and accuracy.

\begin{table}
	\centering
	\caption{Selected Buses for PMU Placement with SFS Algorithm}
	\label{tbl:pmu}
	\begin{tabular}{r|l}
		\hline
		PMU Count & PMU Positions (Bus No.) \\\hline
		1 & 33 \\
		3 & 3, 14, 33 \\
		\textbf{5} & 3, 8, 14, 33, 35 \\
		10 & 1, 3, 4, 8, 10, 14, 25, 33, 34, 35 \\
		15 & 1, 3, 4, 8, 10, 14, 16, 22, 25, 27, 31, 33, 34, 35, 36 \\
		\hline
	\end{tabular}
\end{table}

\begin{table}
	\centering
	\caption{Selected Buses for PMU Placement with SBS Algorithm}
	\label{tbl:pmu_rev}
	\begin{tabular}{r|l}
		\hline
		PMU Count & PMU Positions (Bus No.) \\\hline
		1 & 36 \\
		4 & 1, 17, 35, 36 \\
		\textbf{7} & 1, 8, 17, 21, 35, 36 \\
		8 & 1, 8, 9, 17, 21, 35, 36 \\
		10 & 1, 5, 8, 9, 17, 21, 25, 27, 35, 36 \\
		15 & 1, 2, 5, 8, 9, 10, 17, 21, 23, 25, 27, 28, 35, 36, 38 \\
		\hline
	\end{tabular}
\end{table}

\subsection{PMU Placement Analysis}

In TSAS, the real-time voltage magnitudes and angles of all buses in the power system are employed to assess the post-contingency stability.
Such information requires a large number of high-speed synchrophasors, like from PMUs, to maintain a full observability.
However, such a large-scale installation may not be economically feasible due to high installation cost of PMUs.
Thus it becomes essential to generate a collection of the most effective PMU installation positions in the system.
This can be achieved by analyzing the sensitivity of the measurements from different components in the power system contributing to TSA accuracy and response time.
In this paper we employ the sequential feature selection algorithm \cite{Devijver1982} for this analysis.
Both sequential forward selection (SFS) and sequential backward selection (SBS) are adopted to verify the simulation results, which are demonstrated as follows.

\begin{figure}
	\centering
	\includegraphics[width=\linewidth]{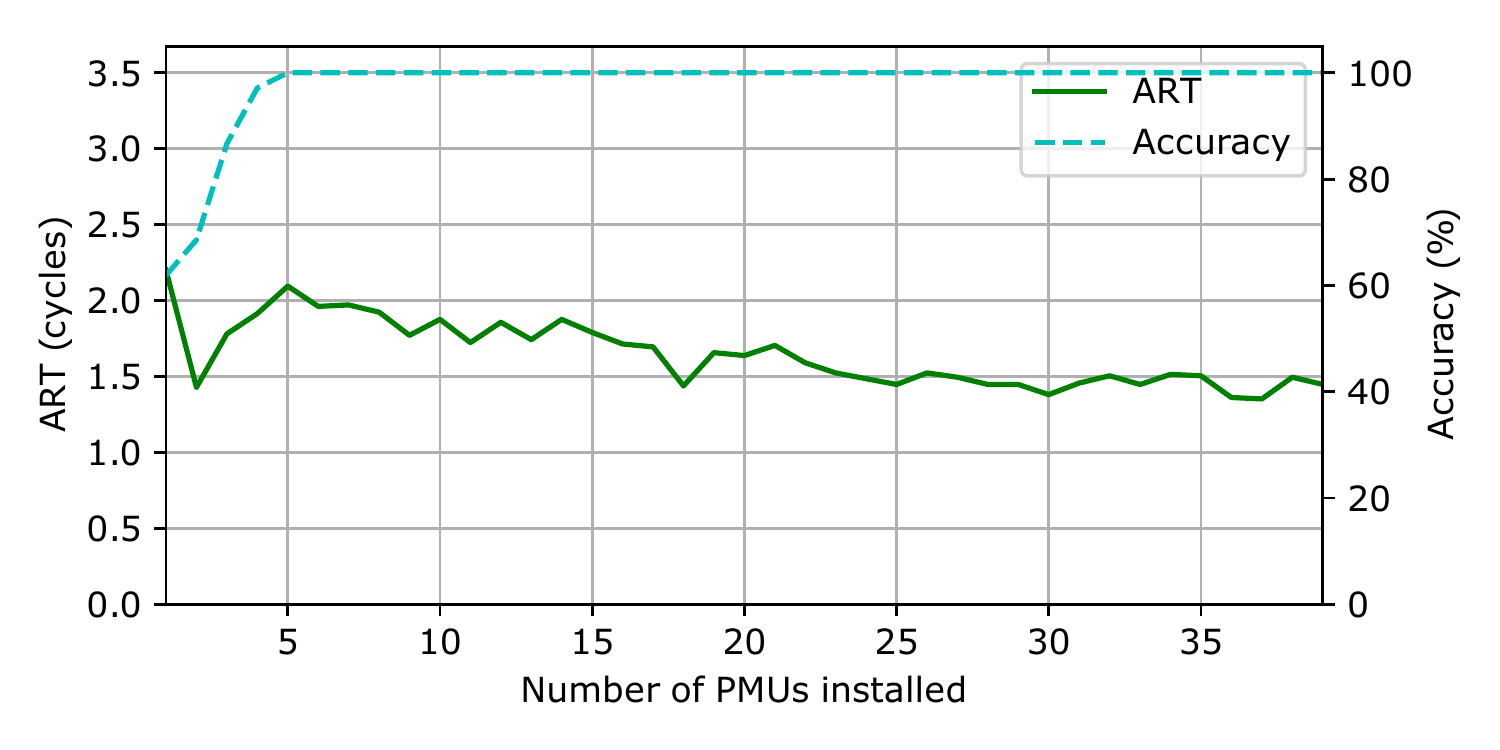}
	\caption{PMU data sensitivity analysis with SFS algorithm.}
	\label{fig:pmu}
\end{figure}

\begin{figure}[ht]
	\centering
	\includegraphics[width=\linewidth]{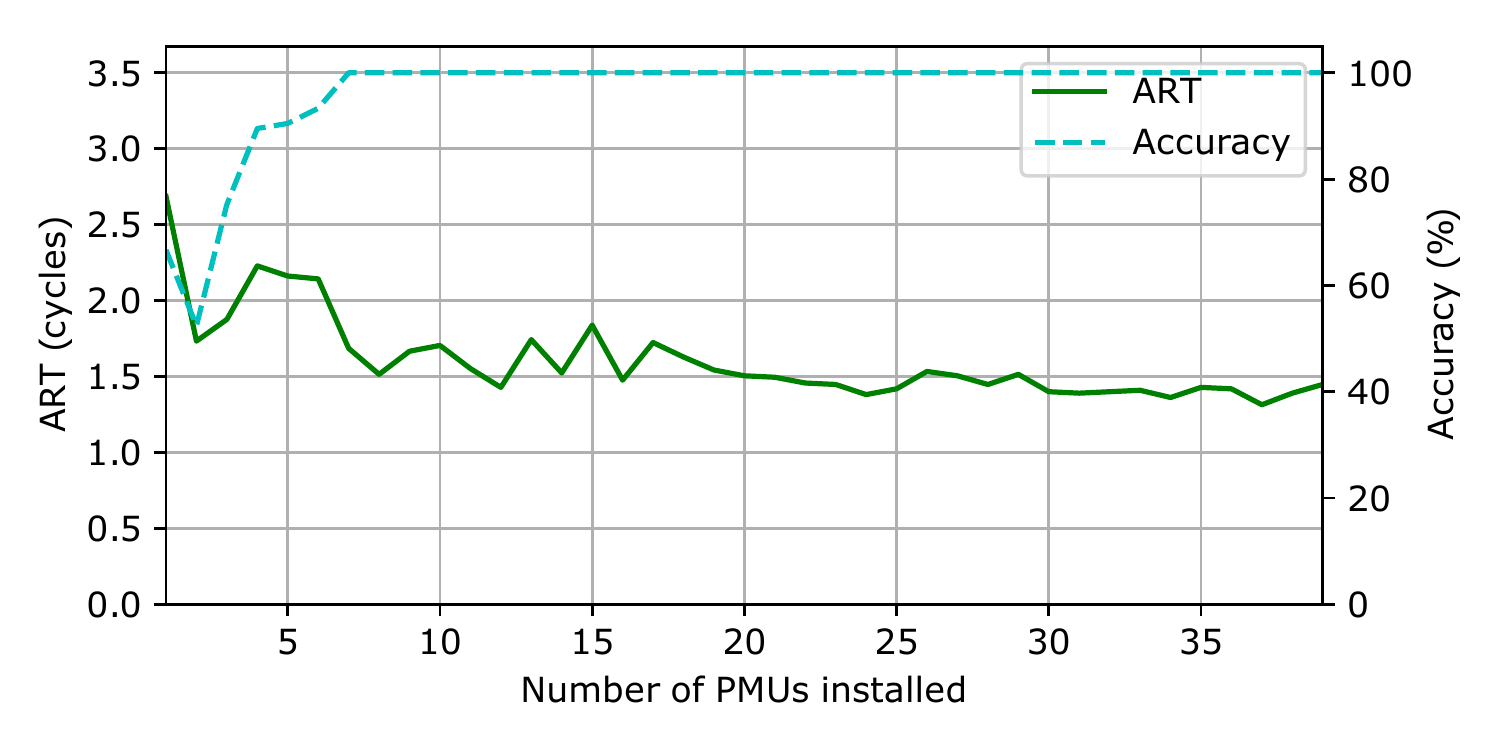}
	\caption{PMU data sensitivity analysis with SBS algorithm.}
	\label{fig:pmu_rev}
\end{figure}

\begin{table*}
	\centering
	\caption{TSA Performance with Noisy PMU Data}
	\label{tbl:10-machine-noisy}
	\begin{tabular}{r|rrrr|rrrr}
	\hline
	\multirow{2}*{$t$} & \multicolumn{4}{c|}{Training Set} & \multicolumn{4}{c}{Testing Set} \\
	\cline{2-9}
	& Unknown & Correct & Wrong & Accuracy & Unknown & Correct & Wrong & Accuracy \\
	\hline
	0 & 3750 & 0 & 0 & N/A & 1250 & 0 & 0 & N/A \\
	1 & 1126 & 2624 & 0 & 100.00\% & 410 & 840 & 0 & 100.00\% \\
	2 & 192 & 3558 & 0 & 100.00\% & 60 & 1190 & 0 & 100.00\% \\
	3 & 111 & 3639 & 0 & 100.00\% & 47 & 1203 & 0 & 100.00\% \\
	4 & 74 & 3676 & 0 & 100.00\% & 26 & 1224 & 0 & 100.00\% \\
	5 & 56 & 3694 & 0 & 100.00\% & 24 & 1226 & 0 & 100.00\% \\
	6 & 28 & 3722 & 0 & 100.00\% & 15 & 1234 & 1 & 99.92\% \\
	7 & 2 & 3747 & 1 & 99.97\% & 1 & 1248 & 1 & 99.92\% \\
	8 & 0 & 3749 & 1 & 99.97\% & 0 & 1249 & 1 & 99.92\% \\
	\hline
	\end{tabular}
\end{table*}

In the SFS algorithm, a number of TSAS with different number of inputs are constructed.
The algorithm first starts with networks with the voltage phasor measurements of one bus as input, and the best performing bus is retained in the next round of selection where the measurements of one more bus are considered as input.
Here the performance comparison is conducted with respect to the corresponding assessment accuracy and ART.
This process repeats until the measurements of all buses are included in the input.
The SBS algorithm employs a similar logic as SFS.
The difference is that SBS starts with the measurements of all buses, and these buses are removed gradually during the selection process \cite{Devijver1982}.
The simulation results are presented in Fig. \ref{fig:pmu} and Fig. \ref{fig:pmu_rev}, and the PMU positions are listed in Tables \ref{tbl:pmu} and \ref{tbl:pmu_rev}.

Figs. \ref{fig:pmu} and \ref{fig:pmu_rev} show that the response time decreases with the increase of installed PMUs, and the assessment accuracy also increases at the same time.
This observation is significant when the total number of PMUs is small (less than 10).
After this initial performance improvement phase, the accuracy remains at 100\% while the ART performance fluctuates around 1.5 cycles, and demonstrates a slight improvement with both of the algorithms.
This leads to a conclusion that a sufficient number of PMUs need to be employed to provide enough system measurements for reliable assessments.
While additional PMUs can marginally decrease the response time needed for assessments, additional equipment cost is incurred.
Real-world implementation shall consider this trade-off when addressing the PMU placement issue.

\subsection{PMU Measurement Accuracy Sensitivity}\label{sub:noisy}

In previous analyses, similar to \cite{Guo2014, Zhang2015a} and other related research, we assume that PMUs in power systems can accurately sample system variables at a high frequency.
Meanwhile, in practice PMUs may suffer from sampling errors.
According to IEEE Standard for Synchrophasor Data Transfer for Power Systems (C37.118.2-2011) \cite{_ieee_2011}, the total vector error for all PMUs complying with the standard should be less than 1\%.
Therefore, a numerical simulation is carried out to study the influence of noisy PMU samples on the performance of TSAS.

In this test, we follow the approach introduced in \cite{he_online_2013} to generate noisy test cases based on the TSAT time-domain simulated power system variables.
Specifically, a random noisy phasor, which satisfies the requirement proposed by \cite{_ieee_2011}, is imposed on all voltage phasors in the dataset.
The manipulated noisy data is employed for both training and testing of TSAS, and all other configurations are identical to previous tests.

The simulation results are summarized in Table \ref{tbl:10-machine-noisy}.
From the table, it can be observed that both ART and accuracy are slightly influenced by the noise, yet the performance decrease is minuscule.
While there is a wrong assessment in the testing cases which was correctly identified with accurate PMU measurements, the training set accuracy remains the same with previous simulation.
Meanwhile, ART for noisy data (1.423 and 1.466 cycles for training and testing cases respectively) is slightly longer than that for noiseless data (1.412 and 1.448 cycles).
To conclude, TSAS can achieve almost the same performance even considering noisy PMU measurements, and can provide superior TSA performance when compared with existing techniques whose results are listed in Table \ref{tbl:comparison}.

\subsection{Large System Test and Computation Time}\label{sub:large}

\begin{table*}
	\centering
	\caption{TSA Performance on 17-generator, 162-bus Test System}
	\label{tbl:17-generator}
	\begin{tabular}{r|rrrr|rrrr}
	\hline
	\multirow{2}*{$t$} & \multicolumn{4}{c|}{Training Set} & \multicolumn{4}{c}{Testing Set} \\
	\cline{2-9}
	& Unknown & Correct & Wrong & Accuracy & Unknown & Correct & Wrong & Accuracy \\
	\hline
	0 & 15000 & 0 & 0 & N/A & 5000 & 0 & 0 & N/A \\
	1 & 4206 & 10794 & 0 & 100.00\% & 1493 & 3507 & 0 & 100.00\% \\
	2 & 3547 & 11453 & 0 & 100.00\% & 1414 & 3586 & 0 & 100.00\% \\
	3 & 2023 & 12977 & 0 & 100.00\% & 876 & 4124 & 0 & 100.00\% \\
	4 & 1198 & 13802 & 0 & 100.00\% & 456 & 4544 & 0 & 100.00\% \\
	5 & 571 & 14429 & 0 & 100.00\% & 214 & 4786 & 0 & 100.00\% \\
	6 & 94 & 14906 & 0 & 100.00\% & 42 & 4958 & 0 & 100.00\% \\
	7 & 12 & 14988 & 0 & 100.00\% & 9 & 4991 & 0 & 100.00\% \\
	8 & 0 & 15000 & 0 & 100.00\% & 0 & 5000 & 0 & 100.00\% \\
	\hline
	\end{tabular}
\end{table*}

\begin{table*}
	\centering
	\caption{TSA Performance on 50-generator, 145-bus Test System}
	\label{tbl:50-generator}
	\begin{tabular}{r|rrrr|rrrr}
	\hline
	\multirow{2}*{$t$} & \multicolumn{4}{c|}{Training Set} & \multicolumn{4}{c}{Testing Set} \\
	\cline{2-9}
	& Unknown & Correct & Wrong & Accuracy & Unknown & Correct & Wrong & Accuracy \\
	\hline
	0 & 15000 & 0 & 0 & N/A & 5000 & 0 & 0 & N/A \\
	1 & 5546 & 9454 & 0 & 100.00\% & 1942 & 3058 & 0 & 100.00\% \\
	2 & 4098 & 10902 & 0 & 100.00\% & 1476 & 3524 & 0 & 100.00\% \\
	3 & 2321 & 12679 & 0 & 100.00\% & 852 & 4148 & 0 & 100.00\% \\
	4 & 1627 & 13373 & 0 & 100.00\% & 639 & 4361 & 0 & 100.00\% \\
	5 & 806 & 14194 & 0 & 100.00\% & 284 & 4716 & 0 & 100.00\% \\
	6 & 70 & 14930 & 0 & 100.00\% & 28 & 4972 & 0 & 100.00\% \\
	7 & 16 & 14984 & 0 & 100.00\% & 6 & 4993 & 1 & 99.98\% \\
	8 & 6 & 14992 & 2 & 99.99\% & 6 & 4993 & 1 & 99.98\% \\
	9 & 4 & 14994 & 2 & 99.99\% & 2 & 4997 & 1 & 99.98\% \\
	10 & 0 & 14997 & 3 & 99.98\% & 0 & 4999 & 1 & 99.98\% \\
	\hline
	\end{tabular}
\end{table*}

Besides the New England 10-machine Test System, we also study two large power system test cases to analyze the scalability of our proposed TSAS, namely a 17-generator, 162-bus system and a 50-generator, 145-bus system \cite{Vittal1992}.
The identical method to generate training and test cases are employed to develop 20 000 contingencies for each test system.
Tests and analyses are performed with the same simulation environment as in previous sections, and the results are presented in Tables \ref{tbl:17-generator} and \ref{tbl:50-generator}.

\begin{figure}[t]
	\centering
	\includegraphics[width=\linewidth]{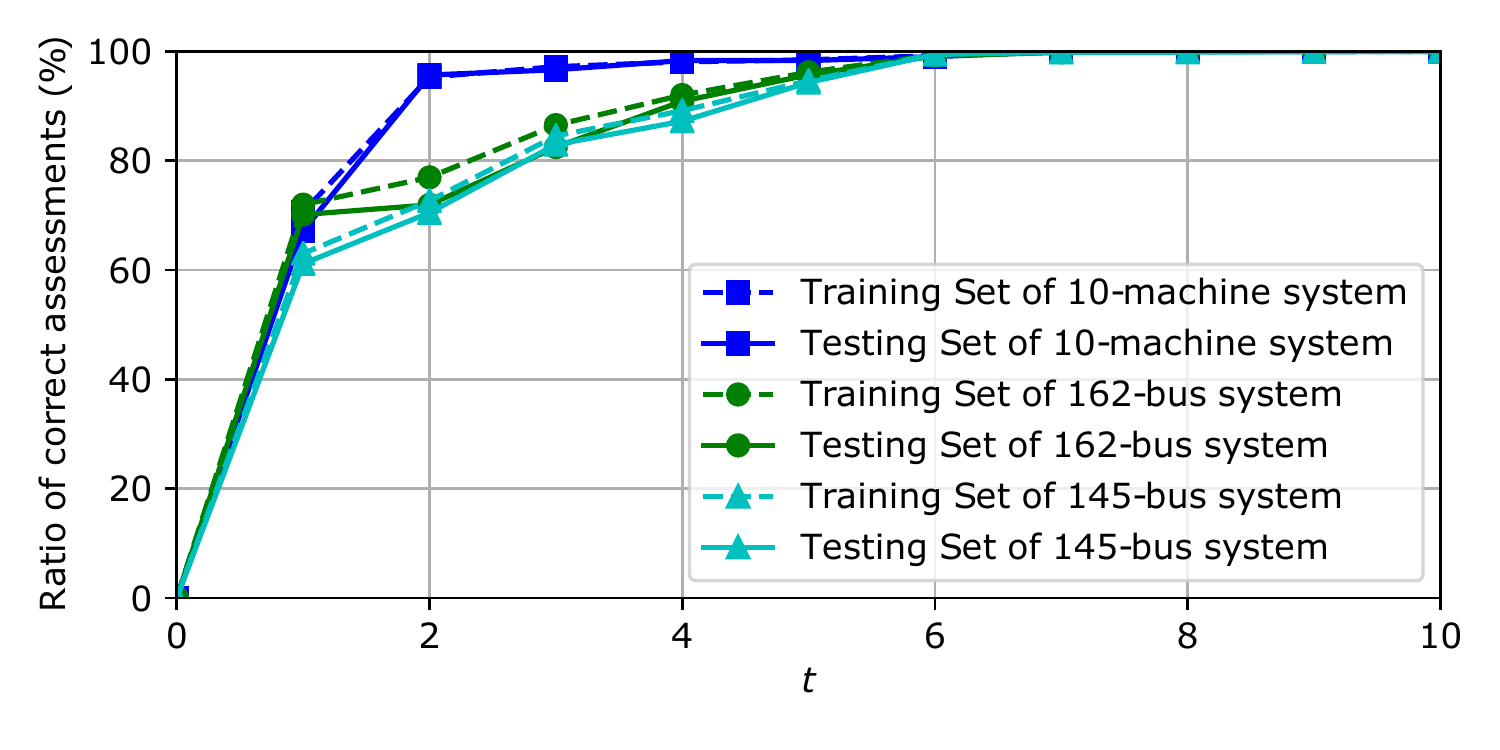}
	\caption{Comparison on the ratio of correct assessment with respect to post-contingency time.}
	\label{fig:ratio}
\end{figure}

From these tables we can see that while the majority of the test cases can still be correctly identified within one cycle only after the clearance of contingencies, the system requires longer ART to perform identifications on both of the larger power grids: 1.901 cycles for the 162-bus system and 2.047 cycles for the 145-bus system.
Meanwhile, the assessment accuracy for both cases remains perfect except for one test case in the 145-bus system.
This observation supports the conclusion that learning from the temporal data-dependencies of the inputs contributes to the improvements in assessment accuracy when compared with the previous state-of-the-art method \cite{Zhang2015a}.

We also depict the response speed of all three tested power systems in Fig. \ref{fig:ratio}.
From this figure it can be observed that the assessment speed is closely related to the system size.
As the New England 10-machine system comprises the least number of generators (10) and buses (39), most assessments can be generated at the very early stage.
Meanwhile, the 162-bus system has more buses than the 145-bus system, but the number of generators are smaller (17 vs. 50).
So they have similar speed.

\begin{table}[t]
	\centering
	\caption{Summary of Training Time}
	\label{tbl:summary}
	\begin{tabular}{l|rr}
		\hline
		Test System & Average & Std. Dev. \\\hline
		New England & 37.43 seconds & 1.62 seconds \\
		162-bus system & 155.86 seconds & 4.75 seconds \\
		145-bus system & 143.16 seconds & 4.69 seconds \\
		\hline
	\end{tabular}
\end{table}

Besides the assessment accuracy and response time, training time is another concern when employing machine learning methods for TSA problems.
Table \ref{tbl:summary} summarizes the training time of the three included tested systems, where each system is trained for 100 times for statistical significance.
From this table, our proposed TSAS is computationally efficient during training.
The training time is roughly linear with the number of training cases, and this number is closely related to the system size.

The fast training speed also makes TSAS robust under significant changes to the power grid.
The system can also be updated online with new measured system variables for performance improvement.
When the control center detects power system topology changes, or schedules periodical TSAS updates, the LSTM system can be efficiently re-trained with existing system dynamics data.
The detailed implementation of TSAS re-training is beyond the scope of this paper and will be investigated in future work.

\section{Conclusion and Future Work}

In this paper we propose a new TSAS for on-line post-contingency stability assessments by utilizing LSTM networks to extract data features.
A time-adaptive scheme is developed to facilitate stability assessments at the earliest possible time instance, reserving more time for the following control actions.
Differently from existing systems, our model learns from the temporal data dependency and utilizes such features to achieve a better assessment accuracy while maintaining the response time performance.
Moreover, our model includes one unique LSTM-based network instead of an ensemble of machine learning components.
This design significantly reduces the model complexity and renders the training process less computationally expensive.

In order to evaluate the performance of our proposed TSAS, three power system test cases are considered, including the New England 10-machine test system, a 17-generator 162-bus system, and a 50-generator 145-bus system.
All simulation results reveal that the proposed model can achieve a superior assessment accuracy within a very short period of time after the clearance of system faults.
Meanwhile, the model is computationally efficient to train using existing measurements, making the model highly scalable for handling TSA tasks in large power systems.

Besides the accuracy and response time tests, we also perform preliminary sensitivity tests on two essential parameters introduced in the proposed system: the stability threshold and training observation window length.
Simulation results suggest a parameter configuration with good performance to facilitate future deployments of the system.
In addition, the sensitivity of PMU measurements is studied, and the sequential feature selection algorithm demonstrates the least number and best locations of PMUs in the tested grid.

Future work will focus on the availability of PMU measurements, considering information loss scenarios and different measurement predictors.
It is assumed in this paper that measurements are available for all PMU installation locations, while a more realistic scenario shall consider cases with missing PMU measurements due to communication delay or loss.
Besides, the predictor employed in this paper is the voltage phasors, while a wider range of such predictors are available for selection \cite{kamwa_development_2009}.
It is interesting to study their sensitivities as system inputs to achieve a better response time.
The online re-training process is another potential future research topic.

\bibliographystyle{IEEEtran}
\bibliography{IEEEabrv,zotero,reference}

\begin{thebibliography}{10}
\providecommand{\url}[1]{#1}
\csname url@samestyle\endcsname
\providecommand{\newblock}{\relax}
\providecommand{\bibinfo}[2]{#2}
\providecommand{\BIBentrySTDinterwordspacing}{\spaceskip=0pt\relax}
\providecommand{\BIBentryALTinterwordstretchfactor}{4}
\providecommand{\BIBentryALTinterwordspacing}{\spaceskip=\fontdimen2\font plus
\BIBentryALTinterwordstretchfactor\fontdimen3\font minus
  \fontdimen4\font\relax}
\providecommand{\BIBforeignlanguage}[2]{{%
\expandafter\ifx\csname l@#1\endcsname\relax
\typeout{** WARNING: IEEEtran.bst: No hyphenation pattern has been}%
\typeout{** loaded for the language `#1'. Using the pattern for}%
\typeout{** the default language instead.}%
\else
\language=\csname l@#1\endcsname
\fi
#2}}
\providecommand{\BIBdecl}{\relax}
\BIBdecl

\bibitem{Kundur2004}
P.~Kundur, J.~Paserba, V.~Ajjarapu, G.~Andersson, A.~Bose, C.~Canizares,
  N.~Hatziargyriou, D.~Hill, A.~Stankovic, C.~Taylor, T.~Van~Cutsem, and
  V.~Vittal, ``Definition and classification of power system stability
  {IEEE/CIGRE} joint task force on stability terms and definitions,''
  \emph{{IEEE} Trans. Power Syst.}, vol.~19, no.~3, pp. 1387--1401, Aug 2004.

\bibitem{chiang_direct_2011}
H.-D. Chiang, \emph{\BIBforeignlanguage{en}{Direct {{Methods}} for {{Stability
  Analysis}} of {{Electric Power Systems}}: {{Theoretical Foundation}}, {{BCU
  Methodologies}}, and {{Applications}}}}.\hskip 1em plus 0.5em minus
  0.4em\relax {John Wiley \& Sons}, Mar. 2011.

\bibitem{fouad_power_1992}
A.-A.~A. Fouad and V.~Vittal, \emph{\BIBforeignlanguage{en}{Power {{System
  Transient Stability Analysis Using}} the {{Transient Energy Function
  Method}}}}.\hskip 1em plus 0.5em minus 0.4em\relax {Prentice Hall}, 1992.

\bibitem{Hiskens1989}
I.~Hiskens and D.~Hill, ``Energy functions, transient stability and voltage
  behaviour in power systems with nonlinear loads,'' \emph{{IEEE} Trans. Power
  Syst.}, vol.~4, no.~4, pp. 1525--1533, Nov. 1989.

\bibitem{xue_quantitative_1999}
Y.~Xue, \emph{Quantitative {{Study}} of {{General Motion Stability}} and an
  {{Example}} on {{Power System Stability}}}.\hskip 1em plus 0.5em minus
  0.4em\relax Nanjing, China: {Jiangsu Science and Technology Press}, 1999, in
  Chinese.

\bibitem{kundur_power_1994}
P.~Kundur, \emph{\BIBforeignlanguage{en}{Power {{System Stability}} and
  {{Control}}}}.\hskip 1em plus 0.5em minus 0.4em\relax {McGraw-Hill
  Education}, Jan. 1994.

\bibitem{Hashiesh2012}
F.~Hashiesh, H.~E. Mostafa, A.-R. Khatib, I.~Helal, and M.~M. Mansour, ``An
  intelligent wide area synchrophasor based system for predicting and
  mitigating transient instabilities,'' \emph{{IEEE} Trans. Smart Grid},
  vol.~3, no.~2, pp. 645--652, Jun. 2012.

\bibitem{Liu1995}
C.~Liu and J.~Thorp, ``Application of synchronised phasor measurements to
  real-time transient stability prediction,'' \emph{IET Generation,
  Transmission and Distribution}, vol. 142, no.~4, pp. 355--360, Jul. 1995.

\bibitem{Pavella2000}
M.~Pavella, D.~Ernst, and D.~Ruiz-Vega, \emph{Transient stability of power
  systems: a unified approach to assessment and control}.\hskip 1em plus 0.5em
  minus 0.4em\relax Norwell: Kluwer Academic Publishers, 2000.

\bibitem{gurusinghe_post-disturbance_2016}
D.~R. Gurusinghe and A.~D. Rajapakse, ``Post-{{Disturbance Transient Stability
  Status Prediction Using Synchrophasor Measurements}},'' \emph{IEEE Trans.
  Power Syst.}, vol.~31, no.~5, pp. 3656--3664, Sep. 2016.

\bibitem{Guo2014}
T.~Guo and J.~Milanovic, ``Probabilistic framework for assessing the accuracy
  of data mining tool for online prediction of transient stability,''
  \emph{{IEEE} Trans. Power Syst.}, vol.~29, no.~1, pp. 377--385, Jan. 2014.

\bibitem{geeganage_application_2015}
J.~Geeganage, U.~D. Annakkage, T.~Weekes, and B.~A. Archer, ``Application of
  {{Energy}}-{{Based Power System Features}} for {{Dynamic Security
  Assessment}},'' \emph{IEEE Trans. Power Syst.}, vol.~30, no.~4, pp.
  1957--1965, Jul. 2015.

\bibitem{Kamwa2012}
I.~Kamwa, S.~Samantaray, and G.~Joos, ``On the accuracy versus transparency
  trade-off of data-mining models for fast-response pmu-based catastrophe
  predictors,'' \emph{{IEEE} Trans. Smart Grid}, vol.~3, no.~1, pp. 152--161,
  Mar. 2012.

\bibitem{wehenkel_automatic_1998}
L.~A. Wehenkel, \emph{Automatic {{Learning Techniques}} in {{Power
  Systems}}}.\hskip 1em plus 0.5em minus 0.4em\relax Norwell, MA, USA: {Kluwer
  Academic Publishers}, 1998.

\bibitem{wehenkel_artificial_1989}
L.~Wehenkel, T.~V. Cutsem, and M.~Ribbens-Pavella, ``An artificial intelligence
  framework for online transient stability assessment of power systems,''
  \emph{IEEE Trans. Power Syst.}, vol.~4, no.~2, pp. 789--800, May 1989.

\bibitem{Zhang2015a}
R.~Zhang, Y.~Xu, Z.~Y. Dong, and K.~P. Wong, ``Post-disturbance transient
  stability assessment of power systems by a self-adaptive intelligent
  system,'' \emph{IET Generation, Transmission and Distribution}, vol.~9,
  no.~3, pp. 296--305, Feb. 2015.

\bibitem{Hochreiter1997}
S.~Hochreiter and J.~Schmidhuber, ``Long short-term memory,'' \emph{Neural
  Computation}, vol.~9, no.~8, pp. 1735--1780, Nov. 1997.

\bibitem{begovic_wide-area_2005}
M.~Begovic, D.~Novosel, D.~Karlsson, C.~Henville, and G.~Michel, ``Wide-{{Area
  Protection}} and {{Emergency Control}},'' \emph{Proc. IEEE}, vol.~93, no.~5,
  pp. 876--891, May 2005.

\bibitem{Pai1989}
A.~Pai, \emph{Energy Function Analysis for Power System Stability}.\hskip 1em
  plus 0.5em minus 0.4em\relax Boston: Kluwer Academic Publishers, 1989.

\bibitem{Narendra1990}
K.~Narendra and K.~Parthasarathy, ``Identification and control of dynamical
  systems using neural networks,'' \emph{{IEEE} Trans. Neural Netw.}, vol.~1,
  no.~1, pp. 4--27, Mar. 1990.

\bibitem{Bengio1994}
Y.~Bengio, P.~Simard, and P.~Frasconi, ``Learning long-term dependencies with
  gradient descent is difficult,'' \emph{{IEEE} Trans. Neural Netw.}, vol.~5,
  no.~2, pp. 157--166, Mar. 1994.

\bibitem{Hochreiter1991}
S.~Hochreiter, ``Untersuchungen zu dynamischen neuronalen netzen,'' Ph.D.
  dissertation, Institut f. Informatik, Technische Univ. Munich, 1991.

\bibitem{UnderstandingLSTMNetworks}
\BIBentryALTinterwordspacing
Understanding {LSTM} networks. [Online]. Available:
  \url{http://colah.github.io/posts/2015-08-Understanding-LSTMs/}
\BIBentrySTDinterwordspacing

\bibitem{lecun_deep_2015}
Y.~LeCun, Y.~Bengio, and G.~Hinton, ``\BIBforeignlanguage{en}{Deep learning},''
  \emph{\BIBforeignlanguage{en}{Nature}}, vol. 521, no. 7553, pp. 436--444, May
  2015.

\bibitem{gomez_support_2011}
F.~Gomez, A.~Rajapakse, U.~Annakkage, and I.~Fernando, ``Support {{Vector
  Machine}}-{{Based Algorithm}} for {{Post}}-{{Fault Transient Stability Status
  Prediction Using Synchronized Measurements}},'' \emph{IEEE Trans. Power
  Syst.}, vol.~26, no.~3, pp. 1474--1483, Aug. 2011.

\bibitem{DiederikKingma2015}
J.~B. Diederik~Kingma, ``Adam: A method for stochastic optimization,'' in
  \emph{Proc. International Conference for Learning Representations}, San
  Diego, CA, U.S., May 2015, pp. 1--15.

\bibitem{hansen_neural_1990}
L.~K. Hansen and P.~Salamon, ``Neural network ensembles,'' \emph{IEEE Trans.
  Pattern Anal. Mach. Intell.}, vol.~12, no.~10, pp. 993--1001, Oct. 1990.

\bibitem{narendra_identification_1990}
K.~S. Narendra and K.~Parthasarathy, ``Identification and control of dynamical
  systems using neural networks,'' \emph{IEEE Trans. Neural Netw.}, vol.~1,
  no.~1, pp. 4--27, Mar. 1990.

\bibitem{Theano}
\BIBentryALTinterwordspacing
Theano documentation. [Online]. Available:
  \url{http://deeplearning.net/software/theano/}
\BIBentrySTDinterwordspacing

\bibitem{Keras}
\BIBentryALTinterwordspacing
Keras: Theano-based deep learning library. [Online]. Available:
  \url{http://keras.io/}
\BIBentrySTDinterwordspacing

\bibitem{Xu2011}
Y.~Xu, Z.~Y. Dong, K.~Meng, R.~Zhang, and K.~P. Wong, ``Real-time transient
  stability assessment model using extreme learning machine,'' \emph{IET
  Generation, Transmission and Distribution}, vol.~5, no.~3, pp. 314--322, Mar.
  2011.

\bibitem{Zhu2015}
L.~Zhu, C.~Lu, and Y.~Sun, ``Time series shapelet classification based online
  short-term voltage stability assessment,'' \emph{{IEEE} Trans. Power Syst.},
  in press.

\bibitem{DSATools}
\BIBentryALTinterwordspacing
Dynamic security assessment software. [Online]. Available:
  \url{http://www.dsatools.com/}
\BIBentrySTDinterwordspacing

\bibitem{Rajapakse2010}
A.~Rajapakse, F.~Gomez, K.~Nanayakkara, P.~Crossley, and V.~Terzija, ``Rotor
  angle instability prediction using post-disturbance voltage trajectories,''
  \emph{{IEEE} Trans. Power Syst.}, vol.~25, no.~2, pp. 947--956, May 2010.

\bibitem{kamwa_development_2009}
I.~Kamwa, S.~Samantaray, and G.~Joos, ``Development of {{Rule}}-{{Based
  Classifiers}} for {{Rapid Stability Assessment}} of {{Wide}}-{{Area
  Post}}-{{Disturbance Records}},'' \emph{IEEE Trans. Power Syst.}, vol.~24,
  no.~1, pp. 258--270, Feb. 2009.

\bibitem{Devijver1982}
P.~Devijver and J.~Kitler, \emph{Pattern Recognition: A Statistical
  Approach}.\hskip 1em plus 0.5em minus 0.4em\relax NJ: Prentice-Hall, 1982.

\bibitem{_ieee_2011}
``{{IEEE Standard}} for {{Synchrophasor Data Transfer}} for {{Power
  Systems}},'' IEEE Std C37.118.2-2011, Dec. 2011.

\bibitem{he_online_2013}
M.~He, V.~Vittal, and J.~Zhang, ``Online dynamic security assessment with
  missing pmu measurements: {{A}} data mining approach,'' \emph{IEEE Trans.
  Power Syst.}, vol.~28, no.~2, pp. 1969--1977, May 2013.

\bibitem{Vittal1992}
V.~Vittal, ``Transient stability test systems for direct stability methods,''
  \emph{{IEEE} Trans. Power Syst.}, vol.~7, no.~1, pp. 37--43, Feb. 1992.

\end{thebibliography}
\end{document}